\newcommand{\bq}{\begin{equation}}
\newcommand{\eq}{\end{equation}}
\newcommand{\bqa}{\begin{eqnarray}}
\newcommand{\eqa}{\end{eqnarray}}

\newcommand{\tv}[2]{\left(\begin{tabular}{c} $#1$ \\ $#2$ \end{tabular}\right)}
\newcommand{\f}{\varphi}
\newcommand{\h}{\frac{1}{2}}

\newcommand{\inti}{\int_{-\infty}^{\infty}}

\documentclass[a4paper, 12pt]{article}

\usepackage{epsfig}
\usepackage{graphicx}
\usepackage{amsmath}
\usepackage{slashed}
\usepackage{makeidx}
\usepackage[top=1in, bottom=1in, left=1in, right=1in]{geometry}
\usepackage{cite}
\usepackage{amsfonts}

\begin{document}

\begin{center}
\LARGE
Ontological States in Non-Interacting Quantum Field Theories\\[30pt]
\large
M.Th.M. van Kessel\footnote{MThMvanKessel@gmail.com}\\[5pt]
July 17, 2024\\[50pt]
\end{center}

\section{Abstract}

This is a paper in the field of ontological deterministic theories behind Quantum Field Theories, like for example the cellular automaton theories proposed by 't Hooft. In these theories one has ontological states in which the state of reality is exactly known and no uncertainties are present. Also these states evolve in time deterministically. A first step in finding the ontological deterministic theory behind the Standard Model is to find in Quantum Field Theory the states that behave as ontological states. We present the ontological states for all non-interacting (3+1-dimensional) Quantum Field Theories occurring in the Standard Model. We summarize the ontological states for free scalar bosons and for free masless Dirac fermions, which are known from the literature. We construct the ontological states for vector bosons, in analogy to the scalar boson case. With this we have a set of ontological states for all particles that are known to occur in reality and in the Standard Model.

\newpage

\section{Introduction}

We believe that behind each Quantum Field Theory (QFT), including of course the Standard Model of elementary particles, there exists an ontological deterministic theory that explains exactly the same physics.

The adjective `ontological' means that, in contrast to common quantum mechanical interpretations, for every state in a QFT, one must be able to say what is actually happening in such a state. This can be either in the form of a certainty, i.e.\ one says that the beables have a certain value, or in the form of probabilities, i.e.\ one says what beable values can be true with which probability. In other words one must be able to say what is the ontology of any state in a QFT.

The adjective `deterministic' means that the reality behind each QFT state evolves in time in a fully deterministic way. If the ontological state of a system is completely known, then it is, at least in principle, also completely known how the system will evolve in time and what the ontological state will be in the future.

Regarding both topics we agree with 't Hooft \cite{CAI} and other pioneers in the field of ontological deterministic theories behind QFTs.

One of the first steps in finding an ontological deterministic theory behind QFT is to find a set of ontological (ontic for short) states in a QFT. These ontic states are QFT states of which we say that all basic properties in the theory (i.e.\ beables) are exactly known, without any probability involved.\footnote{An alternative line of thought here would be to say that for no state in QFT all beables are exactly known. But of course still the states of reality are situations where all beables have a definite value. In this line of thought one would have to explain why the situations where everything is exactly known are not part of the QFT state space. In this paper we shall not follow this line of thought further.} Following the proposal of 't Hooft \cite{CAI} we intend to interprete the other, non-ontic, states in QFT as states where the beables are not exactly known, but with some probability.

More exactly formulated the ontic states are states in the Hilbert space of the QFT that satisfy the following requirements.
\begin{enumerate}
\item Permutation\\
Under time evolution the ontic states evolve according to a permutation. I.e.\ an ontic state shall always evolve into another ontic state of the set (with factor 1 and not any complex factor $e^{i\varphi}$).\\
Clearly this property is needed to have a deterministic theory. If an ontic state would evolve into a state outside of the set, where the beables in the theory are no longer exactly known, then clearly the evolution is not deterministic.
\item Completeness\\
The set of all ontic states spans the full Hilbert space.\\
In the end we have to say, for all states in QFT, what is true in this case or which options are true with which probabilities. If the ontic states do not span the full Hilbert space then there are QFT states for which we have no such explanation in terms of the basic properties of our theory.
\item Othogonality\\
The ontic states are orthogonal.\\
If we wish to say for each ontic state that a certain property is true with probability 1, and if we want to interprete superpositions of ontic states as situations where different options are known with some probabilities, then there can be no overlap between ontic states.\\
Also normalization is needed for the ontic states to be in the QFT state space.
\end{enumerate}

A large amount of work has already been done in finding these ontic states for non-interacting QFTs in 3+1 dimensions. In \cite{CAI}, section 15, 't Hooft presented the ontic states in free masless Dirac theory. In \cite{ontic_bosons} the same author presented ontic states for free (massive) scalar bosons.

In this paper we present a set of ontic states for all non-interacting QFTs that are part of the Standard Model. In section 3 we shall present an overview of the ontic states found for real (uncharged) spin-0 bosons in \cite{ontic_bosons}. In section 4 we do this for complex spin-0 bosons. In section 5 we present an overview of the ontic states for masless spin-$\h$ fermions, found in \cite{CAI} (section 15). In section 6 we add to the list a set of ontic states for vector bosons. These, to the knowledge of the author, have not been presented in literature yet.

With this list the first step in finding an ontological deterministic theory behind the Standard Model has been taken. The next step is to investigate interactions, and see whether these can also be included in such a theory.

Note that in this whole paper we shall work in the Schr\"odinger picture of QFT, because this is best suited to formulate the ontic states.

\section{Real Free Scalar QFT}

The beables, and with it the ontic states, for free scalar QFT for a real field were found in \cite{ontic_bosons}. Here we summarize this result.

The ontic states below will be formulated in terms of the annihilation and creation operators $a$ and $a^{\dagger}$ from scalar QFT. These operators satisfy the wellknown commutation relations
\bqa
\left[ a(\vec{k}), a^\dagger(\vec{l}) \right] &=& (2\pi)^3 \, \omega_k \, \delta^3(\vec{k}-\vec{l}) \nonumber\\
\left[ a(\vec{k}), a(\vec{l}) \right] &=& 0 \nonumber\\
\left[ a^\dagger(\vec{k}), a^\dagger(\vec{l}) \right] &=& 0
\eqa
with $\omega_k \equiv c \sqrt{\vec{k}^2+\mu^2}$, $\mu \equiv (mc)/\hbar$ and $m$ the mass of the scalar particle. With the creation operators we can build a set of basis states for the QFT Hilbert space. These basis states are
\bq \label{basis_states_SQFT}
a^\dagger(\vec{k}_1) \ldots a^\dagger(\vec{k}_n) \, |0\rangle \,,
\eq
where $|0\rangle$ is the vacuum state of the free QFT.

To formulate the ontic states we discretize momentum space on a lattice with spacing $\Delta k$ and enumerate the vectors $\vec{k}$ with a discrete index $i$.

\subsection{The Ontic States}

The ontic states for real free scalar (spin-0) QFT, that are found in \cite{ontic_bosons}, are:
\bq\label{OS_SQFT}
|\varphi_1, \varphi_2, \ldots\rangle_\mathrm{ont} = \prod_i \left[ \sum_{n_i=0}^\infty \frac{1}{\sqrt{2\pi}} \frac{1}{\sqrt{n_i!}} \left( \frac{\Delta k^{3/2}}{(2\pi)^{3/2} \sqrt{c\sqrt{\vec{k}_i^2+\mu^2}}} \right)^{n_i} e^{-i\varphi_i n_i} \left( a^\dagger(\vec{k}_i) \right)^{n_i} \right] |0\rangle
\eq
Here the product over $i$ is a product over all discrete values of $\vec{k}$, i.e.\ all $\vec{k}_i$'s.

Each ontic state is defined by the values $\f_i$, which can take on the values in $[0, 2\pi\rangle$, and there is one such value $\f_i$ for each $\vec{k}_i$. In \cite{ontic_bosons} these values are interpreted as cogwheel positions and there are as many cogwheels as there are momentum values $\vec{k}_i$.

These states are the eigenstates of the beables found in \cite{ontic_bosons} and are just the ontic states for a harmonic oscillator, in the continuum limit. Also here we express these ontic states in terms of the annihilation and creation operators $a$ and $a^\dagger$ of the QFT, which will later allow for a natural extension to the vector boson case.

We can also write the expression for these states in a shorter way:
\bq
|\varphi_1, \varphi_2, \ldots\rangle_\mathrm{ont} \equiv |\f(\vec{k})\rangle_\mathrm{ont}
\eq

Now it can be seen that these states satisfy all requirements (presented in the introduction) for a set of ontic states:

Under time evolution these states permute. If at $t=0$ the state is
\bq
|\psi(0)\rangle = |\varphi_1, \varphi_2, \ldots\rangle_\mathrm{ont} \,,
\eq
then at some other time $t$ the state is
\bqa
|\psi(t)\rangle &=& \prod_i \Bigg[ \sum_{n_i=0}^\infty \frac{1}{\sqrt{2\pi}} \frac{1}{\sqrt{n_i!}} \left( \frac{\Delta k^{3/2}}{(2\pi)^{3/2} \sqrt{c\sqrt{\vec{k}_i^2+\mu^2}}} \right)^{n_i} \nonumber\\
&& \phantom{\prod_i \Bigg[ \sum_{n_i=0}^\infty} e^{-i\varphi_i n_i} \left( a^\dagger(\vec{k}_i) \, e^{-ic\sqrt{\vec{k}_i^2+\mu^2}t} \right)^{n_i} \Bigg] |0\rangle \nonumber\\
&=& |\, \varphi_1 + c\sqrt{\vec{k}_1^2+\mu^2}t \mod 2\pi, \varphi_2 + c\sqrt{\vec{k}_2^2+\mu^2}t \mod 2\pi, \ldots\rangle_\mathrm{ont} \,.
\eqa
Here we recognize the cogwheel evolution law
\bq
\f \rightarrow \f + c\sqrt{\vec{k}^2+\mu^2} \, t \mod 2\pi
\eq
from \cite{ontic_bosons}.

These states span the full Hilbert space. Any state (\ref{basis_states_SQFT}) can be made from the ontic states via
\bq
\frac{1}{2\pi} \int_0^{2\pi} d\varphi_1 \, e^{+i\varphi_1 n_1} \, \frac{1}{2\pi} \int_0^{2\pi} d\varphi_2 \, e^{+i\varphi_2 n_2} \ldots |\varphi_1, \varphi_2, \ldots\rangle_\mathrm{ont} \,.
\eq

And last these states can be shown to be orthogonal:
\bq
_\mathrm{ont}\langle\varphi_1', \varphi_2', \ldots |\varphi_1, \varphi_2, \ldots\rangle_\mathrm{ont} = \delta(\varphi_1-\varphi_1' \mod 2\pi) \, \delta(\varphi_2-\varphi_2' \mod 2\pi) \ldots
\eq

Note that the ontic states we find look a bit like coherent states, which are eigenstates of the annihilation operator $a$. These coherent states are:
\bq
|z, g\rangle = e^{-\h|z|^2} \, \sum_{n=0}^{\infty} \frac{z^n}{n!} \, \left( \inti d^3k \, g(\vec{k}) \, a^\dagger(\vec{k}) \right)^n \, |0\rangle
\eq
One can make these states evolve among themselves, as is required for ontic states, by choosing
\bq
g(\vec{k}) = e^{i\varphi(\vec{k})} \, g'(\vec{k}),
\eq
however then these states do not span the full Hilbert space and are not orthogonal. This could be expected since with only the $a$'s one cannot build Hermitean operators, for which the eigenstates would be a complete set. Therefore the operators $a$ are not proper beables, as is also concluded in \cite{ontic_bosons}.

\subsection{The Beables}

The ontic states (\ref{OS_SQFT}) are eigenstates of the beables in the theory. In this case these beables are exactly the operators $b$ from \cite{ontic_bosons}. This we can show as follows.

The beables belonging to the ontic states (\ref{OS_SQFT}) are
\bqa
\varphi_\mathrm{op}(\vec{k_i}) &=& \int_0^{2\pi} d\varphi_1 \, \int_0^{2\pi} d\varphi_2 \, \ldots \int_0^{2\pi} d\varphi_{i-1} \, \int_0^{2\pi} d\varphi_i \, e^{-i\varphi_i} \, \int_0^{2\pi} d\varphi_{i+1} \, \ldots \nonumber\\\vspace{5pt}
&& \hspace{150pt} |\varphi_1, \varphi_2, \ldots\rangle_\mathrm{ont} \, _\mathrm{ont}\langle\varphi_1, \varphi_2, \ldots | \,. \label{beable_SQFT}
\eqa
Here in principal the eigenvalue $e^{i\varphi_i}$ can be chosen anyway one likes, here we chose it such as to get exactly to the operators $b$ in \cite{ontic_bosons}, eq.\ (4.8).

Now by plugging in the ontic states in terms of the $a^\dagger$'s (\ref{OS_SQFT}) and performing the integrals over the $\varphi$'s one can show that this becomes
\bq
\varphi_\mathrm{op}(\vec{k_i}) = \left(I+\alpha^2 \, a^\dagger(\vec{k}_i) a(\vec{k}_i)\right)^{-\h} \alpha \, a(\vec{k}_i) \,,
\eq
with
\bq \label{alpha}
\alpha = \frac{\Delta k^{3/2}}{(2\pi)^{3/2} \sqrt{c\sqrt{\vec{k}_i^2+\mu^2}}} \,,
\eq
which are the beables $b$ found in \cite{ontic_bosons}, eq.\ (4.8) (apart from the factor $\alpha$). By choosing $e^{+i\varphi_i}$ in (\ref{beable_SQFT}) we would get the other beables $b^\dagger$ from \cite{ontic_bosons}:
\bq
\varphi_\mathrm{op}^\dagger(\vec{k_i}) = \alpha \, a^\dagger(\vec{k}_i) \, \left(I+\alpha^2 \, a^\dagger(\vec{k}_i) a(\vec{k}_i)\right)^{-\h}
\eq

The factor $\alpha$ arises because we expressed the ontic states in terms of the continuum annihilation and creation operators, while in \cite{ontic_bosons} the beables are expressed in terms of the discrete annihilation and creation operators, which satisfy $[a, a^\dagger] = 1$.

\section{Complex Free Scalar QFT}

Having the ontic states (\ref{OS_SQFT}) for a real scalar field, where one has one creation operator $a^\dagger$ and one annihilation operator $a$, it is straightforward to find the ontic states for a complex scalar field, where one has two creation operators $a^\dagger$, $b^\dagger$ and two annihilation operators $a$, $b$. Of course the $a$ ($a^\dagger$) and $b$ ($b^\dagger$) commute, which is important for the construction of ontic states below.

\subsection{The Ontic States}

The ontic states in this case are:
\bqa
&& |\varphi^{(1)}_1, \varphi^{(1)}_2, \ldots, \varphi^{(2)}_1, \varphi^{(2)}_2, \ldots\rangle_\mathrm{ont} = \nonumber\\\nonumber\\
&& |\f^{(1)}(\vec{k}), \f^{(2)}(\vec{k})\rangle_\mathrm{ont} = \nonumber\\\nonumber\\
&& \qquad \prod_{i_1} \left[ \sum_{n_{i_1}=0}^\infty \frac{1}{\sqrt{2\pi}} \frac{1}{\sqrt{n_{i_1}!}} \left( \frac{\Delta k^{3/2}}{(2\pi)^{3/2} \sqrt{c\sqrt{\vec{k}_{i_1}^2+\mu^2}}} \right)^{n_{i_1}} e^{-i\varphi^{(1)}_{i_1} n_{i_1}} \left( a^\dagger(\vec{k}_{i_1}) \right)^{n_{i_1}} \right] \nonumber\\
&& \qquad \prod_{i_2} \left[ \sum_{n_{i_2}=0}^\infty \frac{1}{\sqrt{2\pi}} \frac{1}{\sqrt{n_{i_2}!}} \left( \frac{\Delta k^{3/2}}{(2\pi)^{3/2} \sqrt{c\sqrt{\vec{k}_{i_2}^2+\mu^2}}} \right)^{n_{i_2}} e^{-i\varphi^{(2)}_{i_2} n_{i_2}} \left( b^\dagger(\vec{k}_{i_2}) \right)^{n_{i_2}} \right]|0\rangle \nonumber\\
\label{OS_CSQFT}
\eqa
Again here the product over the $i$'s is a product over all values of the momentum $\vec{k}_i$.

Thus in the case of a complex scalar field the ontic states are described by 2 sets of cogwheel positions, which we label with upper indices as $\f^{(1)}$ and $\f^{(2)}$. Again for each set there is one cogwheel per momentum value $\vec{k}_i$.

That these states satisfy all requirements for ontic states can be verified in the same way as in section 3. Here it is crucial that the $a$ ($a^\dagger$) and $b$ ($b^\dagger$) commute.

\subsection{The Beables}

The beables are also very similar to the ones for a real scalar field, only here we have 2 sets of beables: 
\bqa
\varphi^{(1)}_\mathrm{op}(\vec{k_i}) &=& \left(I+\alpha^2 \, a^\dagger(\vec{k}_i) a(\vec{k}_i)\right)^{-\h} \alpha \, a(\vec{k}_i) \nonumber\\
\varphi^{(2)}_\mathrm{op}(\vec{k_i}) &=& \left(I+\alpha^2 \, b^\dagger(\vec{k}_i) b(\vec{k}_i)\right)^{-\h} \alpha \, b(\vec{k}_i) \nonumber\\
\varphi^{(1)\dagger}_\mathrm{op}(\vec{k_i}) &=& \alpha \, a^\dagger(\vec{k}_i) \, \left(I+\alpha^2 \, a^\dagger(\vec{k}_i) a(\vec{k}_i)\right)^{-\h} \nonumber\\
\varphi^{(2)\dagger}_\mathrm{op}(\vec{k_i}) &=& \alpha \, b^\dagger(\vec{k}_i) \, \left(I+\alpha^2 \, b^\dagger(\vec{k}_i) b(\vec{k}_i)\right)^{-\h}
\eqa
$\alpha$ Is again given by (\ref{alpha}).

\section{Masless Free Dirac QFT}

The beables and ontic states for non-interacting masless Dirac fermions were found in \cite{CAI}, section 15. Here we summarize the results from there.

Note that here, in contrast to bosons, one has to set the mass $m$ to zero, in order for the construction to work.

To formulate the ontic states one first has to choose the Weyl or chiral representation of the $\gamma$-matrices. See e.g.\ \cite{Itzykson}, section 2-1-3:
\bqa
\gamma^0 &=& \left(\begin{tabular}{cccc} 0 & 0 & -1 & 0 \\ 0 & 0 & 0 & -1 \\ -1 & 0 & 0 & 0 \\ 0 & -1 & 0 & 0 \end{tabular}\right) \quad,
\gamma^1 = \left(\begin{tabular}{cccc} 0 & 0 & 0 & 1 \\ 0 & 0 & 1 & 0 \\ 0 & -1 & 0 & 0 \\ -1 & 0 & 0 & 0 \end{tabular}\right) \quad, \nonumber\\
\gamma^2 &=& \left(\begin{tabular}{cccc} 0 & 0 & 0 & -i \\ 0 & 0 & i & 0 \\ 0 & i & 0 & 0 \\ -i & 0 & 0 & 0 \end{tabular}\right) \quad,
\gamma^3 = \left(\begin{tabular}{cccc} 0 & 0 & 1 & 0 \\ 0 & 0 & 0 & -1 \\ -1 & 0 & 0 & 0 \\ 0 & 1 & 0 & 0 \end{tabular}\right)
\eqa

In this representation, and with the fermion mass set to zero, the theory splits up in 2 parts, one part for the upper 2 components of the $\psi$-operator field and one part for the lower 2 components. Below we shall focus on the upper 2-spinor, but the same reasoning can be followed for the lower 2-spinor.

The upper 2-spinor satisfies the Weyl equation:
\bq
\left(\begin{tabular}{cc} $\frac{1}{c} \frac{\partial}{\partial t} + \frac{\partial}{\partial z}$ & $\frac{\partial}{\partial x} - i\frac{\partial}{\partial y}$ \\
$\frac{\partial}{\partial x} + i\frac{\partial}{\partial y}$ & $\frac{1}{c} \frac{\partial}{\partial t} - \frac{\partial}{\partial z}$ \end{tabular}\right) \tv{\psi_1(\vec{x}, t)}{\psi_2(\vec{x}, t)} = \tv{0}{0}
\eq

As described in \cite{CAI}, section 15, the 2-spinor field can be expanded as
\bq \label{expansion_psi}
\psi_\alpha(\vec{x}) = \frac{1}{(2\pi)^3} \int d^3k \, e^{+i\vec{k}\cdot\vec{x}} \left( u^+_\alpha(\vec{k}) \, a_1(\vec{k}) + u^-_\alpha(\vec{k}) \, a_2^\dagger(-\vec{k}) \right)
\eq
with
\bq
u^\pm(\vec{k}) = \frac{1}{\sqrt{2|\vec{k}|(|\vec{k}| \pm k_z)}} \tv{\pm|\vec{k}| + k_z}{k_x + ik_y} \,.
\eq

Now $a_1$ and $a_2$ satisfy the wellknown anti-commutation rules for momentum space annihilation and creation operators:
\bqa
\{ a_i(\vec{k}), a_j(\vec{l})\} &=& 0 \nonumber\\
\{ a_i^\dagger(\vec{k}), a_j^\dagger(\vec{l})\} &=& 0 \nonumber\\
\{ a_i(\vec{k}), a_j^\dagger(\vec{l})\} &=& (2\pi)^3 \, \delta_{ij} \, \delta^3(\vec{k}-\vec{l})
\eqa

\subsection{The Ontic States}

With the momentum space annihilation and creation operators formulated above in (\ref{expansion_psi}) we can now formulate the ontic states. These states were found in \cite{CAI}, section 15. Here we summarize them.

Like in \cite{CAI} we will write the momentum vectors $\vec{k}$ as follows:
\bq
\vec{k} \equiv s \hat{q} |\vec{k}|
\eq
Here the $\hat{q}$, $s$ and $|\vec{k}|$ are defined as follows:
\begin{itemize}
\item $\hat{q}$: unit 3-vector ($|\hat{q}|=1$) with $\hat{q}_z \geq 0$
\item $s$: $s = \pm 1$
\item $|\vec{k}|$: $|\vec{k}| \in [0, +\infty\rangle$
\end{itemize}

Like for scalars above we will discretize these variables and label them with an index $j$, to formulate the ontic states. The lattice discrete volume element for $\hat{q}$ is $\sin\theta\Delta\theta\Delta\f$, where $\theta$ and $\f$ are the polar coordinates for $\hat{q}$.

Following \cite{CAI} we also introduce a variable $r$, $r \in \langle-\infty, +\infty\rangle$. Also this variable we discretize with a lattice spacing $\Delta r$.

We shall enumerate the sets $\{\hat{q}, s, r\}$ by an index $j$: $\{\hat{q}_j, s_j, r_j\}$

Note that the number of possible values for $\{\hat{q}, s, r\}$ is twice as large as the amount of possible values for $\vec{k} \equiv s \hat{q} |\vec{k}|$, because the $r$ ranges over $\langle-\infty, +\infty\rangle$ while $|\vec{k}|$ ranges over $[0, +\infty\rangle$.

Now the ontic states are:
\bqa
|n_1, n_2, \ldots\rangle_\mathrm{ont} &=& \prod_j \Big[ \sqrt{\sin\theta\Delta\theta\Delta\varphi\Delta r} \frac{1}{(2\pi)^2} \int_0^\infty d|\vec{k}_j| \, |\vec{k}_j| \Big( a_1^\dagger(s_j\hat{q}_j|\vec{k}_j|) \, e^{-is_j|\vec{k}_j|r_j} + \nonumber\\
&& \phantom{\prod_j \Big[ \sqrt{\sin\theta\Delta\theta\Delta\varphi\Delta r} \frac{1}{(2\pi)^2} \int_0^\infty d|\vec{k}_j| \, |\vec{k}_j| \Big(} a_2(s_j\hat{q}_j|\vec{k}_j|) \, e^{+is_j|\vec{k}_j|r_j} \Big) \Big]^{n_j} \nonumber\\
&& \qquad \prod_i \left( \frac{\Delta k^{3/2}}{(2\pi)^{3/2}} \, a_2^\dagger(s_i\hat{q}_i|\vec{k}_i|) \, e^{+ic|\vec{k}_j|t} \right) |0\rangle \label{OS_DQFT}
\eqa
Here the product over $j$ is a product over all values $\{\hat{q}_j, s_j, r_j\}$ and the product over $i$ is a product over all values $\vec{k}_i$.

Also here we can write the ontic states shorter as
\bq
|n_1, n_2, \ldots\rangle_\mathrm{ont} \equiv |n(\hat{q}, s, r)\rangle_\mathrm{ont} \,.
\eq

Note the complex exponential factor $e^{+ic|\vec{k}_j|t}$ in the expression (\ref{OS_DQFT}). This means that for each time $t$ we define the ontic states slightly differently to make sure that the ontic states evolve into another ontic state, i.s.o.\ into another ontic state multiplied by a complex value of modulus 1. This is allowed because in QM one can always multiply all states by the same $e^{i\f}$.

Note also that the overall sign of this expression depends on the order in which the product is taken, because of the anti-commutation relations for the $a$'s. We will not specify this order, any order is correct, as long as it is used consistently in all expressions.

We see that in the case of Dirac fermions the ontic states are described by a value 0 or 1 per value of $\{\hat{q}_j, s_j, r_j\}$.

For the ontic states (\ref{OS_DQFT}) one can check that these satisfy all requirements for ontic states:

If at $t=0$ the state is
\bq
|\psi(0)\rangle = |n_1, n_2, \ldots\rangle_\mathrm{ont} = |n(\hat{q}, s, r)\rangle_\mathrm{ont}
\eq
then at time $t$ the state is
\bqa
|\psi(t)\rangle &=& \prod_j \Big[ \sqrt{\sin\theta\Delta\theta\Delta\varphi\Delta r} \frac{1}{(2\pi)^2} \int_0^\infty d|\vec{k}_j| \, |\vec{k}_j| \Big( a_1^\dagger(s_j\hat{q}_j|\vec{k}_j|) \, e^{-is_j|\vec{k}_j|(r_j+s_jct)} + \nonumber\\
&& \phantom{\prod_j \Big[ \sqrt{\sin\theta\Delta\theta\Delta\varphi\Delta r} \frac{1}{(2\pi)^2} \int_0^\infty d|\vec{k}_j| \, |\vec{k}_j| \Big(} a_2(s_j\hat{q}_j|\vec{k}_j|) \, e^{+is_j|\vec{k}_j|(r_j+s_jct)} \Big) \Big]^{n_j} \nonumber\\
&& \qquad \prod_i \left( \frac{\Delta k^{3/2}}{(2\pi)^{3/2}} \, a_2^\dagger(s_i\hat{q}_i|\vec{k}_i|) \right) |0\rangle \,.
\eqa
If now we choose the order of the product such that $s=\pm 1$ are clearly separated, e.g.\
\bq
\prod_j = \prod_{\hat{q}} \prod_s \prod_r
\eq
then the order is conserved under time evolution, and no additional minus signs arise from re-ordering into the original order. Then we have
\bq
|\psi(t)\rangle = |n(\hat{q}, s, r-sct)\rangle_\mathrm{ont} \,.
\eq
In other words the $n$'s are permuted according to
\bq
\{\hat{q}, s, r\} \rightarrow \{\hat{q}, s, r + sct\} \,.
\eq

Also every state in Dirac QFT,
\bq
a_1^\dagger(\vec{k}_1) \ldots a_1^\dagger(\vec{k}_m) a_2^\dagger(\vec{l}_1) \ldots a_2^\dagger(\vec{k}_n) |0\rangle = \prod_{i_1} \left( a_1^\dagger(\vec{k}_{i_1}) \right)^{n_{i_1}} \prod_{i_2} \left( a_2^\dagger(\vec{l}_{i_2}) \right)^{n_{i_2}} |0\rangle
\eq
can be expressed in terms of the states $|n_1, n_2, \ldots\rangle_\mathrm{ont}$. This can most easily be done by expressing the $a$'s and $|0\rangle$ in terms of the ontic operators $\psi(\hat{q}, s, r)$ and the ontic empty state $|0, 0, \ldots\rangle$ as defined in \cite{CAI}, section 15.

This means that the ontic states (\ref{OS_DQFT}) also span the full Hilbert space.

Last they can be shown to be orthogonal:
\bq
_\mathrm{ont}\langle n'_1, n'_2, \ldots|n_1, n_2, \ldots\rangle_\mathrm{ont} = \prod_i \delta_{n'_i, n_i}
\eq
Here we used the anti-commutation relation
\bqa
&& \Bigg\{ \sqrt{\sin\theta\Delta\theta\Delta\varphi\Delta r} \frac{1}{(2\pi)^2} \int_0^\infty d|\vec{k}_i| \, |\vec{k}_i| \Big( a_1^\dagger(s_i\hat{q}_i|\vec{k}_i|) \, e^{-is_i|\vec{k}_i|r_i} + a_2(s_i\hat{q}_i|\vec{k}_i|) e^{+is_i|\vec{k}_i|r_i} \Big) \,, \nonumber\\
&& \phantom{\Bigg\{} \sqrt{\sin\theta\Delta\theta\Delta\varphi\Delta r} \frac{1}{(2\pi)^2} \int_0^\infty d|\vec{k}_j| \, |\vec{k}_j| \Big( a_1^\dagger(s_j\hat{q}_j|\vec{k}_j|) \, e^{-is_j|\vec{k}_j|r_j} + a_2(s_j\hat{q}_j|\vec{k}_j|) e^{+is_j|\vec{k}_j|r_j} \Big) \Bigg\} \nonumber\\
&& \qquad = \delta_{ij} \nonumber\\\label{rel_DQFT}
\eqa

\subsection{The Beables}

With the ontic states (\ref{OS_DQFT}) above the beables are
\bqa
n_\mathrm{op}(\hat{q}_j, s_j, r_j) &=& \sum_{n_1=0}^1 \, \sum_{n_2=0}^1 \, \ldots \sum_{n_{j-1}=0}^1 \, \sum_{n_j=0}^1 \, n(\hat{q}_j, s_j, r_j) \, \sum_{n_{j+1}=0}^1 \, \ldots \nonumber\\\vspace{5pt}
&& \hspace{150pt} |n_1, n_2, \ldots\rangle_\mathrm{ont} \, _\mathrm{ont}\langle n_1, n_2, \ldots | \,.
\eqa
Plugging in the ontic states (\ref{OS_DQFT}), working out the sums over the $n$'s and using the anti-commutation relations, one then finds
\bqa
n_\mathrm{op}(\hat{q}_j, s_j, r_j) &=& \sin\theta\Delta\theta\Delta\varphi\Delta r \frac{1}{(2\pi)^2} \int_0^\infty d|\vec{k}_j| \, |\vec{k}_j| \Big( a_1^\dagger(s_j\hat{q}_j|\vec{k}_j|) \, e^{-is_j|\vec{k}_j|r_j} + \nonumber\\
&& \phantom{\sin\theta\Delta\theta\Delta\varphi\Delta r \frac{1}{(2\pi)^2} \int_0^\infty d|\vec{k}_j| \, |\vec{k}_j| \Big(} a_2(s_j\hat{q}_j|\vec{k}_j|) e^{+is_j|\vec{k}_j|r_j} \Big) \nonumber\\
&& \phantom{\sin\theta\Delta\theta\Delta\varphi\Delta r} \frac{1}{(2\pi)^2} \int_0^\infty d|\vec{k}_j| \, |\vec{k}_j| \Big( a_1(s_j\hat{q}_j|\vec{k}_j|) \, e^{+is_j|\vec{k}_j|r_j} + \nonumber\\
&& \phantom{\sin\theta\Delta\theta\Delta\varphi\Delta r \frac{1}{(2\pi)^2} \int_0^\infty d|\vec{k}_j| \, |\vec{k}_j| \Big(} a_2^\dagger(s_j\hat{q}_j|\vec{k}_j|) e^{-is_j|\vec{k}_j|r_j} \Big)
\eqa
Introducing the ontic field $\psi(\hat{q}, s, r)$ as in \cite{CAI}, section 15,
\bq
\psi(\hat{q}, s, r) = \frac{1}{(2\pi)^2} \int_0^\infty d|\vec{k}| \, |\vec{k}| \Big( a_1(s\hat{q}|\vec{k}|) \, e^{+is|\vec{k}|r} + a_2^\dagger(s\hat{q}|\vec{k}|) e^{-is|\vec{k}|r} \Big)
\eq
this becomes:
\bq
n_\mathrm{op}(\hat{q}_j, s_j, r_j) = \beta \, \psi^\dagger(\hat{q}_j, s_j, r_j) \, \psi(\hat{q}_j, s_j, r_j) \,,
\eq
with
\bq
\beta \equiv \sin\theta\Delta\theta\Delta\varphi\Delta r \,,
\eq
which are exactly the beables as found in \cite{CAI}, section 15.

\section{Vector Boson QFT}

To the author's knowledge there are no ontic states known for free spin-1 bosons in the literature. However, knowing the ontic states for scalar bosons (\ref{OS_SQFT}), one can build very similar ontic states for vector bosons.

Also for vector boson QFTs one has annihilation and creation operators, only now one has more than 1.

For a real massive vector field one has 3 $a$'s and $a^\dagger$'s that satisfy
\bqa
\left[ a^{(\lambda)}(\vec{k}), a^{(\lambda)}(\vec{l}) \right] &=& 0 \nonumber\\
\left[ a^{(\lambda)\dagger}(\vec{k}), a^{(\lambda)\dagger}(\vec{l}) \right] &=& 0 \nonumber\\
\left[ a^{(\lambda)}(\vec{k}), a^{(\lambda)\dagger}(\vec{l}) \right] &=& (2\pi)^3 \, \omega_k \, \delta_{\lambda\lambda'} \, \delta^3(\vec{k}-\vec{l}) \,,\label{CR_VB}
\eqa
with $\lambda = 1, 2, 3$ and $\omega_k \equiv c \sqrt{\vec{k}^2+\mu^2}$. See e.g.\ \cite{Itzykson}, section 3-2-3.

For a massless vector boson gauge symmetries are important and some special steps are required to arrive at the annihilation and creation operators. One can for example follow the approach in \cite{Jauch}, section 6-2. There one arrives at the following operators.
\bqa
\left[ a_r(\vec{k}), a^\dagger_s(\vec{l}) \right] &=& (2\pi)^3 \, \omega_k \, \zeta_s \, \delta_{rs} \, \delta^3(\vec{k}-\vec{l}) \nonumber\\
\left[ a_r(\vec{k}), a_s(\vec{l}) \right] &=& 0 \nonumber\\
\left[ a^\dagger_r(\vec{k}), a^\dagger_s(\vec{l}) \right] &=& 0
\eqa
Here $r$ ans $s$ can take on the values 0, \ldots, 3, $\omega_k$ is defined as
\bq
\omega_k \equiv c |\vec{k}| \,,
\eq
and $\zeta_s$ is defined as
\bq
\zeta_s \equiv \left\{ \begin{tabular}{ll}
-1  & if $s = 0$ \\
1  & if $s = 1, 2, 3$
\end{tabular} \right.
\eq
In this formalism $a_0$ is a creation operator, in contrast to what the absence of the dagger suggests. Together with these commutation relations come subsidiary conditions on the state space, which forbid certain states.

Despite these difficulties however one can create a set of ontic states, using the creation operators $a^\dagger_1$, $a^\dagger_2$, $a^\dagger_3$ and $a_0$. Also in the ontological theory part of the state space is now not allowed because of the subsidiary conditions, which comes from the gauge invariance. It will be no problem to forbid a certain part of the state space and with it certain ontic states, since these parts of the Hilbert space are clearly separated, as is also required in QFT. For example a state that is forbidden will never evolve into a state that is allowed by the subsidiary conditions.

Other approaches to deal with massless vector bosons also exist, e.g.\ the Gupta-Blueler formalism which is described in \cite{Jauch}, section 6-3, or in \cite{Schweber}. Also here one can apply the reasoning below to construct ontic states.

\subsection{The Ontic States}

Continuing with massive vector bosons, for which there are 3 $a$'s and $a^\dagger$'s which satisfy (\ref{CR_VB}), we find the following ontic states.
\bqa
&& |\varphi^{(1)}_1, \varphi^{(1)}_2, \ldots, \varphi^{(2)}_1, \varphi^{(2)}_2, \ldots\varphi^{(3)}_1, \varphi^{(3)}_2, \ldots\rangle_\mathrm{ont} = \nonumber\\\nonumber\\
&& |\f^{(1)}(\vec{k}), \f^{(2)}(\vec{k}), \f^{(3)}(\vec{k})\rangle_\mathrm{ont} = \nonumber\\\nonumber\\
&& \qquad \prod_{i_1} \left[ \sum_{n_{i_1}=0}^\infty \frac{1}{\sqrt{2\pi}} \frac{1}{\sqrt{n_{i_1}!}} \left( \frac{\Delta k^{3/2}}{(2\pi)^{3/2} \sqrt{c\sqrt{\vec{k}_{i_1}^2+\mu^2}}} \right)^{n_{i_1}} e^{-i\varphi^{(1)}_{i_1} n_{i_1}} \left( a^{(1)\dagger}(\vec{k}_{i_1}) \right)^{n_{i_1}} \right] \nonumber\\
&& \qquad \prod_{i_2} \left[ \sum_{n_{i_2}=0}^\infty \frac{1}{\sqrt{2\pi}} \frac{1}{\sqrt{n_{i_2}!}} \left( \frac{\Delta k^{3/2}}{(2\pi)^{3/2} \sqrt{c\sqrt{\vec{k}_{i_2}^2+\mu^2}}} \right)^{n_{i_2}} e^{-i\varphi^{(2)}_{i_2} n_{i_2}} \left( a^{(2)\dagger}(\vec{k}_{i_2}) \right)^{n_{i_2}} \right] \nonumber\\
&& \qquad \prod_{i_3} \left[ \sum_{n_{i_3}=0}^\infty \frac{1}{\sqrt{2\pi}} \frac{1}{\sqrt{n_{i_3}!}} \left( \frac{\Delta k^{3/2}}{(2\pi)^{3/2} \sqrt{c\sqrt{\vec{k}_{i_3}^2+\mu^2}}} \right)^{n_{i_3}} e^{-i\varphi^{(3)}_{i_3} n_{i_3}} \left( a^{(3)\dagger}(\vec{k}_{i_3}) \right)^{n_{i_3}} \right]|0\rangle \nonumber\\
\label{OS_VQFT}
\eqa
Again here the product over the $i$'s is a product over all values of the momentum $\vec{k}_i$.

Thus in the case of a massive vector field the ontic states are described by 3 sets of cogwheel positions, which we label with upper indices as $\f^{(1)}$, $\f^{(2)}$ and $\f^{(3)}$. Again for each set there is one cogwheel per momentum value $\vec{k}_i$.

That these states satisfy all requirements for ontic states can be verified in the same way as in section 3.

For complex fields one can proceed in the same way and one gets to 6 sets of cogwheels.

\subsection{The Beables}

Here the beables can be found in the same way as for scalar QFT. In the case of real massive vector boson QFT the beables are:
\bqa
\varphi^{(\lambda)}_\mathrm{op}(\vec{k_i}) &=& \left(I + \alpha^2 \, a^{(\lambda)\dagger}(\vec{k}_i) a^{(\lambda)}(\vec{k}_i)\right)^{-\h} \alpha \, a^{(\lambda})(\vec{k}_i) \nonumber\\
\varphi^{(\lambda)\dagger}_\mathrm{op}(\vec{k_i}) &=& \alpha \, a^{(\lambda)\dagger}(\vec{k}_i) \, \left(I + \alpha^2 \, a^{(\lambda)\dagger}(\vec{k}_i) a^{(\lambda)}(\vec{k}_i)\right)^{-\h}
\eqa
Here $\lambda = 1, 2, 3$ and $\alpha$ is given in (\ref{alpha}).

\section{Discussion}

To the authors knowledge the sets of ontic states presented above are the only ones known for the 3+1-dimensional QFTs discussed. It is not known whether these are the only existing sets or that there are more sets.

Do note that for spin-0 bosons a construction like for spin-1/2 Dirac fermions to make ontic states, like (\ref{OS_DQFT}), doesn't work, because of the commutation (i.s.o.\ anti-commutation) relations for bosons. Due to these commutation relations a relation like (\ref{rel_DQFT}) doesn't hold and the states can't be shown to span the full Hilbert space and shown to be orthogonal.

Note also that for spin-$\h$ fermions a construction like for spin-0 bosons, (\ref{OS_SQFT}), would fail. Because of the anti-commutation relations for fermions the product over $i$ there would only include the values $n_i=0$ and $n_i=1$. This would mean that for fermions we would have that the would-be ontic states satisfy the constraints
\bq
\frac{1}{2\pi} \int_0^{2\pi} d\varphi_i \, e^{+i\varphi_i n_i} \, |\varphi_1, \varphi_2, \ldots\rangle_\mathrm{ont} = 0 \quad \textrm{if } n_i \geq 2 \,, \textrm{for all } i \,.
\eq

Note that for lower dimensions other sets of ontic states are known, e.g.\ \cite{Wetterich}.

Note also that for bosons (scalar and vector) ontic states are known for the free theory with mass $\mu$ general (also $\mu\neq 0$), while for Dirac fermions only for $\mu=0$. We suspect that for fermions the mass term has to be introduced via interaction, with the Higgs field.

\section{Outlook}

Above we found a set of ontic states for all types of particles present in the Standard Model (for the non-interacting case). The idea is now that for these states we know exactly what is going on in reality. Thus, e.g.\ for scalar bosons, we say that, in case the quantum state $|\psi\rangle$ is equal to one of the ontic states (\ref{OS_SQFT}), reality is exactly known and all cogwheel positions $\f$ (one value per momentum value $\vec{k}_i$) are exactly known.

If now we are in any other quantum state, which is a superposition of the ontic states (this is always the case due to the second requirement for ontic states), following the ideas in \cite{CAI}, we say that reality can be in a number of situations, each with their own probability. These probabilities are related to the coefficients in the superposition, as explained in \cite{CAI}. The evolution of this situation in time is completely dictated by the evolution of the ontic states, i.e.\ probability in = probability out. Now because the ontic states evolve according to the Schr\"odinger equation, and because this equation is linear, the superpositions also satisfy the usual QFT evolution in time.

The next big question is now how exactly the typical quantum effects arise. E.g.\ how do we explain quantum interference? And how can we explain the quantum 2-slit experiment results?

For this we have to explain, from the ontological deterministic theory, why certain coefficients in our superposition become important. For example, for a measurement of the charge observable, QM says that after the measurement the system will be in an eigenstate of this observable, and the probabilities for each eigenstate are dictated by the coefficients of these eigenstates in the original superposition. Thus also from the ontological deterministic theory we have to explain why, just after the measurement, the system gets into an eigenstate of the observable. Also we have to explain why the probability for this is related to the coefficient of this eigenstate in the original superposition.

In other words, we have to explain why the eigenstates of observables, which are typically complex superpositions of the ontic states, are relevant for measurement outcomes. Clearly, because all measurements are interactions, here the interactions will become crucial. Therefore the most important next step in this research is to find how to include interactions in an ontological deterministic theory. Only when we know exactly what is happening at the ontological level during an interaction (needed for the measurement) we will be able to understand why the system evolves into one of the eigenstates during the measurement.

A lot of work has already been done on this topic, see e.g.\ \cite{Hooft_int} and \cite{Hooft_int_2}. The next step is to find explicitly the interactions at the ontological level with which we can explain the interactions known in Nature, i.e.\ the ones in the Standard Model.

\end{document}